\documentclass[english,twocolumn,twocolumn,showpacs,preprintnumbers,amsmath,amssymb,superscriptaddress]{revtex4}
\usepackage[T1]{fontenc}
\usepackage[latin1]{inputenc}
\usepackage{graphicx}

\makeatletter

\makeatletter
\makeatletter

\makeatletter
\makeatletter
\makeatletter

\makeatletter

\makeatletter

\makeatletter

\makeatletter

\makeatletter

\@ifundefined{definecolor}
 {\@ifundefined{definecolor}
 {\@ifundefined{definecolor}
 {\usepackage{color}

}{}
}{}
}{}

\usepackage{dcolumn}

\usepackage{bm}

\def\lsim{\mathrel{\rlap{\lower4pt\hbox{\hskip1pt$\sim$}}
    \raise1pt\hbox{$<$}}}         %less than or approx. symbol
\def\gsim{\mathrel{\rlap{\lower4pt\hbox{\hskip1pt$\sim$}}
    \raise1pt\hbox{$>$}}}         %greater than or approx. symbol

\makeatother

\makeatother

\makeatother

\makeatother

\def\citet{\cite}

\makeatother

\makeatother

\makeatother

\usepackage{babel}
\makeatother
\begin{document}

\title{ Landau hydrodynamical model at RHIC and LHC }

\author{Sheng-Xu Liu}

\author{Hui-Jie Wang}

\author{Fu-Ming Liu}

\email{liufm@iopp.ccnu.edu.cn}

\address{Key laboratory of Quark and Lepton Physics (MoE) and Institute of
Particle Physics, Central China Normal University, Wuhan 430079, China}

\date{\today}

\begin{abstract}
The rapidity distribution and transverse spectra of most copious particles such as pions, Kaons and antiprotons from central Au+Au collisions at $\sqrt{s_{NN}}=200$~GeV and central Pb+Pb collisions at $\sqrt{s_{NN}}=2.76$~TeV have been investigated in the framework of Landau hydrodynamical model. 
With a more realistic choice of freeze-out condition and the employment of lattice equation of state, we find transverse expansion of the collision systems is important to explain the observed data. With the increase of collision energy from RHIC to LHC, transverse flow becomes more and more important for hadron production at midrapidity, especially for more massive particle.   
\end{abstract}
\maketitle
%\pacs{25.75.Nq,24.10.Nz}
%\keywords{}

\section{Introduction}

The Relativistic Heavy Ion collider (RHIC) and the Large Hadronic
Collider (LHC), the biggest high energy nuclear experiments in the
world, realized the creation of a quark gluon plasma (QGP), a thermalized
system consist of free quarks and gluons, which was the main matter
of our early universe shortly after the Big Bang. In order to understand
the organ of our universe and to study the non-perturbative region
of quantum chromodynamics, it is of great importance to measure the
macroscopic properties of this matter, such as equation of state,
transport coefficient, viscosity, phase boundary, and so on. However,
direct measurement is extremely difficult because the QGP matter made-in-lab
exists only shortly in violent systems, due to the confinement of
quarks. In fact, those properties can only be extracted from experimental
data based phenomenological models.

Among those models, statistical thermal model was first proposed in
1950's~\cite {Fermi}, though there was neither the terminology of
quark nor that of QGP at that time. This model is still in use today.
The basic assumption is, after the creation of a fireball (of QGP),
all hadron species are strongly produced under the same temperature 
followed with weak decays and hadronic interactions. The phase transition 
temperature $T_{c}$  is one of the main parameters of the model, which can
be extracted from experimental data of particle yields/ratios. The
obtained temperature is about 160~MeV, very close to Lattice (LQCD) prediction\cite{Aoki:2009sc}.

However, statistical thermal model can neither describe the dynamics
evolution of the collision systems nor cover more observables than
particle yields/ratios. Soon Laudau~\citet{Landau} proposed hydrodynamical
model, which made one step forward. With latter development, more
hydrodynamical models have been employed in heavy
ion physics, from Bjorken's one-dimensional boost-invariant hydrodynamics to hydrodynamics in two-dimensional and three-dimensional space, additonally with viscosity and fluctuated initialization. Thus hydrodynamics has become one of the most powerful models, with more and more experimental observables explained, with a more and more realistic but also complicate hydrodynamical evolution. In this paper we will
develop the recent work\citet{Wong,Csernai}, investigate the employment of the olddest and simplest hydrodynamical model, Landau hydrodynamical model, to RHIC and LHC experiments.

The paper is organized as following. In section II, we will introduce
our calculation approach of particle production based on Landau model.
In section III we will present our results on both rapidity distribution
and transverse spectra of the most copious particles such as charged
pions, Kaons and antiproton from central Au+Au collisions at $\sqrt{s_{NN}}=200$~GeV
and central Pb+Pb collisions at $\sqrt{s_{NN}}=2.76$~TeV, and compare
with RHIC and LHC data. The effects from different freeze-out conditions
and equation of states will be shown as well. In
section IV we will draw the conclusions on what we have learned from
this exercise.

\section{Calculation approach}

Hydrodynamics is often employed to describe the evolution of the collision
system, from an initial time (initial condition) to a freeze-out time
(FO condition). During this period, the system is governed with the
energy-momentum conservation law \begin{eqnarray}
\partial_{\mu}T^{\mu\nu}=0.\end{eqnarray}
 Solving this fluid equation with a given dynamical equation of state
(EOS), and decomposing the energy-momentum tensor \begin{eqnarray}
T^{\mu\nu}=(\epsilon+P)u^{\mu}u^{\nu}-Pg^{\mu\nu},\label{T^munu}\end{eqnarray}
 one can obtain the energy density $\epsilon$, pressure $P$, flow
velocity $u^{\nu}$ at each space-time point of the collision system.
Here we assume an ideal hydrodynamics for the system. More term should
be included in (\ref{T^munu}) once viscosity is considered.

The proper time coordinate $\tau$, space-time rapidity coordinate
$\eta$, $r$ and $\phi$ are often used to note each space-time point.
Thus \begin{eqnarray}
 & x= & r\cos\phi,\nonumber \\
 & y= & r\sin\phi,\nonumber \\
 & z= & \tau\sinh\eta,\nonumber \\
 & t= & \tau\cosh\eta.\label{xyzt}\end{eqnarray}
 Or inversely \begin{equation}
\eta=\frac{1}{2}\ln\frac{t+z}{t-z}.\label{eq:rapi field}\end{equation}

The initial state of the Landau hydrodynamical model is a thin static
disk of thickness $\Delta$ and diameter $a$. It is assumed to be
the approximation of the overlap of two highly Lorentz contracted
nuclei, and the two geometrical sizes are assumed to be related by
\begin{equation}
\Delta=a/\gamma,\end{equation}
 with $\gamma$ the Lorentz factor corresponding to the velocity of
the colliding nuclei in the center-of-mass frame, \begin{equation}
\gamma=\frac{\sqrt{S_{NN}}}{2m_{N}},\end{equation}
 where $S_{NN}$ is the invariant collision energy per colliding nucleons
of mass $m_{N}$.

The Landau hydrodynamical evolution of the system is divided in two
parts: the initial longitudinal expansion (during which transverse
velocities and displacements are neglected), and the subsequent \char`\"{}conic
flight\char`\"{}, where transverse velocities appear and hadronic
particles freeze out of the fluid.

Assuming a simple dynamical EOS of the form \begin{equation}
P=\epsilon/3,\label{eq:EOS}\end{equation}
 an approximate solution of the 1+1 dimensional problem of the longitudinal
expansion phase is given by Landau~\citet{Landau}. This solution
is summarized (using the notation of \citet{Wong}) as follow. The
energy-density field is \begin{equation}
\epsilon=\epsilon_{0}\exp\left\{ -\frac{4}{3}\left(y_{+}+y_{-}-\sqrt{y_{+}y_{-}}\right)\right\} ,\label{eq:e field}\end{equation}
 where $y_{\pm}$ are logarithmic light-cone coordinates \begin{equation}
y_{\pm}=\ln\frac{t\pm z}{\Delta},\label{eq:ypm}\end{equation}
 and $\epsilon_{0}$ is the initial energy-density of the disk. The
flow four-velocity are expressed as \begin{equation}
u^{\mu}=(\cosh{\eta},0,0,\sinh{\eta}).\label{eq:u y}\end{equation}
 The momentum rapidity of the flow \begin{equation}
y_{{\rm flow}}=\frac{1}{2}\ln\frac{u_{0}+u_{z}}{u_{0}-u_{z}}\end{equation}
 coincides with the space-time rapidity coordinate $\eta$ defined in Eq.(\ref{eq:rapi field}),
similarly to the case of the Bjorken model.

Now is the step of \char`\"{}conic flight\char`\"{}, with hadronic
particles freezing out of the fluid. The four-momentum of each particle
can be specified as \begin{equation}
p^{\mu}=(m_{T}\cosh{y},p_{T}\cos{\phi_{p}},p_{T}\sin{\phi_{p}},m_{T}\sinh{y})\label{eq:p mu}\end{equation}
 in terms of its transverse momentum $p_{T}$, rapidity $y$ and azimuth
angle $\phi_{p}$, where $m_{T}=\sqrt{m^{2}+p_{T}^{2}}$ is the transverse
mass, with the mass of the particle, $m$.

The distribution of emitted hadrons in the momentum space is calculated
with the Cooper-Frye formula~\citet{Cooper_Frye} \begin{equation}
\frac{d^{3}N}{dyd^{2}p_{T}}=\int f(x,p)p^{\mu}d\sigma_{\mu},\label{eq:CF}\end{equation}
 where $f(x,p)$ is the phase-space distribution of particles at freeze-out.
The phase-space distribution has the thermal form \begin{equation}
f(x,p)=\tilde{f}\left(\frac{p^{\mu}u_{\mu}(x)}{T(x)}\right)=\frac{g}{(2\pi)^{3}}\frac{1}{\exp\left(\frac{p^{\mu}u_{\mu}(x)}{T(x)}\right)+\xi},\label{eq:ps dist}\end{equation}
 with $\xi$=1 for fermions, -1 for bosons and 0 for classical particles
obeying Boltzmann statistics. $g$ is a degeneracy factor that is
different for each particles species. The temperature at a given space-time
point ${T(x)}$ is obtained from the energy-density field in eq.(\ref{eq:e field})
with a certain EOS.

\begin{figure}
\includegraphics[scale=0.9]{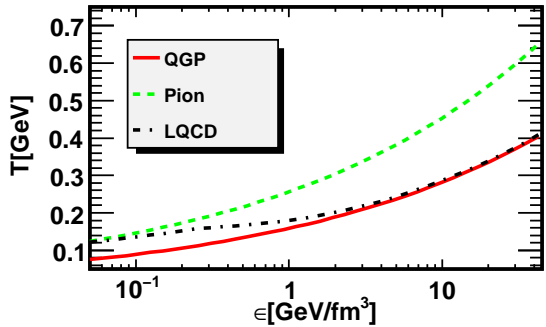}

\caption{\label{fig:Fig1} (Color Online) Temperature dependence of energy
density. Dashed line: calculated according to Pion gas; Solid line:
calculated with QGP freedom; Dotted-dashed line: LQCD result\cite{Aoki:2005vt}. }
\end{figure}

In previous work~\citet{Csernai}, a hadronic gas EOS has been used,
where the massive particles in the hadronic phase is assumed to obey
the relativistic Boltzmann distribution. Then the pressure of the
medium is given by \begin{equation}
P(T)=\sum_{\alpha}\frac{g_{\alpha}}{\pi^{2}}m_{\alpha}^{2}T^{2}K_{2}(m_{\alpha}/T),\end{equation}
 where $m_{\alpha}$ and $g_{\alpha}$ are the mass and degeneracy
factor of particle type $\alpha$, and $K_{2}$ is a modified Bessel
function of the second kind. The energy-density and temperature is
related via \begin{equation}
\epsilon(T)=3\sum_{\alpha}P_{\alpha}\left(1+\frac{m_{\alpha}}{3T}\frac{K_{1\alpha}}{K_{2\alpha}}\right),\label{eq:edens T}\end{equation}
 where we used the notation \begin{equation}
K_{i\alpha}=K_{i}(m_{\alpha}/T).\end{equation}
 In Fig.~\ref{fig:Fig1}, this relation is plotted with dashed line
and note as \char`\"{}Pion\char`\"{}, since the lightest mesons contribute
the most to the pressure and energy density of hadronic gas. The same
relation calculated with quark and gluon degree of freedom has been
plotted with solid line and noted as \char`\"{}QGP\char`\"{}. We also
plot Lattice result\cite{Aoki:2005vt} as dotted dashed line and note as \char`\"{}LQCD\char`\"{}.
One can see Lattice EOS coincides with QGP one at high energy density
and coincides with Pion one at low energy density. It is clear that
no pions at very high energy density and no QGP at very low temperature,
and LQCD result is a more realistic choice. In the following we repeat
the previous calculation with Pion EOS and also take LQCD EOS as a
comparison.

\begin{figure*}
\includegraphics[scale=0.8]{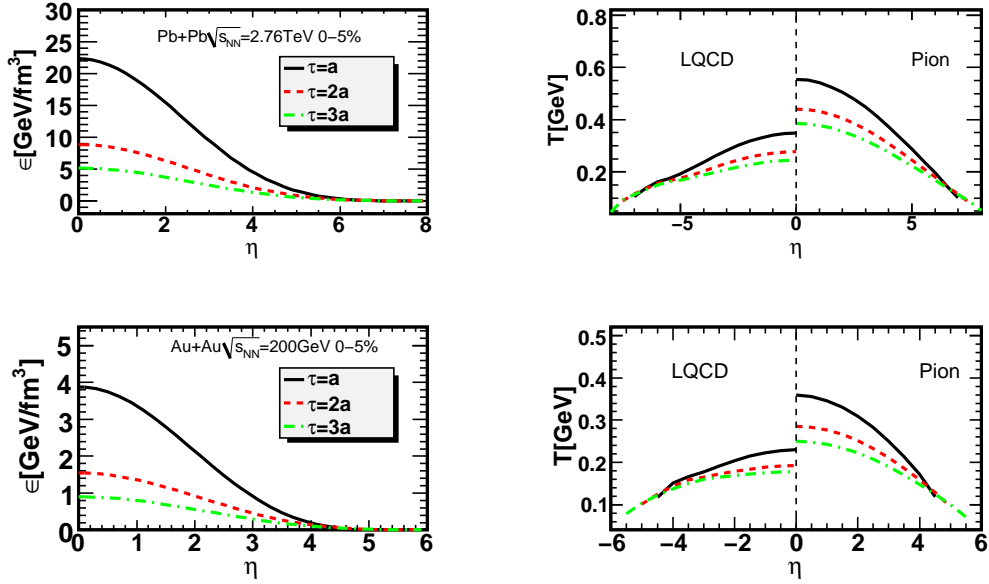}

\caption{\label{fig:Fig2} (Color Online) Left panel:energy density dependence
of $\eta$ at different proper time $\tau$ where upper panel for central Pb+Pb collision at $\sqrt{s_{NN}}=2.76$~TeV and lower
panel for the central Au+Au collision at $\sqrt{s_{NN}}=200$~GeV. Right
panel: the corresponding temperature as a function of $\eta$ and $\tau$
with two choices of EOS. }
\end{figure*}

We will employ Landau fluid dynamics in to RHIC and LHC heavy ion
collisions, in the most simple case, central collisions. Therefore,
the Landau solution of energy density in eq.(\ref{eq:e field}) is
plotted as a function of $\eta$ and $\tau$ in left panel of Fig.~\ref{fig:Fig2},
where upper panel for the central Pb+Pb collision at $\sqrt{s_{NN}}=2.76$~TeV
and lower panel for the central Au+Au collision at $\sqrt{s_{NN}}=200$~GeV.
The corresponding temperature is plotted in the right panel, based
on two choices of EOS.

To calculate hadrons' spectrum with Cooper-Frye formula in eq.(\ref{eq:CF}),
we treat the hypersurface element four-vector \begin{equation}
d\sigma_{\mu}=(dxdydz,dydzdt,dxdzdt,dxdydt)\label{eq:dsigma mu1}\end{equation}
 at each space-time point of FO surface, in terms of $\tau,\eta,r$
and $\phi$ according to eq.(\ref{xyzt}) for the convenience of calculation.

Insert $u^{\mu}u_{\mu}=1$ into eq.(\ref{eq:CF}), and make use of
eq.(\ref{eq:u y}), one can get \begin{equation}
\label{pmu_umu}
p^{\mu}u_{\mu}=m_{T}\cosh(y-\eta)\end{equation}
 and \begin{equation}
\label{umu_sigmamu}
u^{\mu}d\sigma_{\mu}\rightarrow\frac{a^{2}\pi}{4}\tau_{{\rm FO}}d\eta,\end{equation}
 where the integral over $r$ and $\phi$ has been done.

In Refs.~\citet{Landau,Wong} an estimate of the time of freeze-out
is \begin{equation}
\tau_{{\rm FO}}=2a.\label{eq:TR cond}\end{equation}
 We take this as a choice of FO condition. Currently the popular  FO condition is at a fixed temperature around $T_{c}\sim160$~MeV. Here
we take a constant energy density $\epsilon_{{\rm FO}}$ as another
choice of FO condition, which corresponds to a fixed temperature via
EOS. 
Since FO is a change of the treatment on collision systems, 
from a collective mode to the collection of many single particles,
this change should happen after the appearance of hadrons.
Therefore, a realistic FO condition, should not occur at too high
energy density where the matter is still in QGP phase. 
According to Fig.~1, three values 0.08, 0.16 and 0.24~GeV/fm$^{3}$ were employed for 
$\epsilon_{{\rm FO}}$, because Pion EOS and LQCD EOS coincide at $\epsilon \sim 0.08$~GeV/fm$^{3}$  and 0.24~GeV/fm$^{3}$ roughly corresponds to $T_C \sim 160MeV$ according to LQCD EOS.

With the FO condition \begin{equation}
\epsilon=\epsilon_{{\rm FO}}\end{equation}
 and the solution of Landau fluid in eq.(\ref{eq:e field}), we can
get the freeze-out time at each given space point as the following
form, \begin{equation}
\tau_{{\rm FO}}=\Delta\exp\frac{4B+\sqrt{4B^{2}-12\eta^{2}}}{6},\end{equation}
 where the constant $B=\frac{3}{4}\ln(\epsilon_{0}/\epsilon_{{\rm FO}})$,
$\epsilon_{0}$ the initial energy-density of the disk, $\Delta$
the initial thickness of the disk and $\eta$ the space-time rapidity
of the given space point. Then the integral along freez-out hypersurface can be done easily with Eqs.(\ref{pmu_umu}, \ref{umu_sigmamu}).

\section{Results and discussion}

We first repeated the previous work~\citet{Csernai}, calculated
the rapidity spectra of charged pions from central Au+Au collision
at $\sqrt{s_{NN}}=200$~GeV with the same FO condition $\tau_{{\rm FO}}=2a$
and the same EOS. The result is shown as a dashed line in Fig.~3,
a very nice coincidence with BRAHMS data where empty squares for $\pi^{+}$
and empty cycles for $\pi^{-}$. However, this nice coincidence is just an accidence. 
We explain this in the following.

In Fig.~3, the dashed line was decomposed into several dotted lines which are the contributions from different $\eta$ bin [-6,-4],[-4,-2], ..., [4,6], respectively.
The energy density and the temperature along $\eta$-axis  at  $\tau_{\rm FO}=2a$ correspond to the second lines on the lower panel of Fig.~2.
As one can see, most midrapidity pions come out from  the fluid at  $\eta \sim 0$   where the temperature  is as high as 300~MeV. 
This is not realistic because hadrons should not appear at such a high temperature, where the baryonic chemical potentil is almost zero. In fact, this energy density is too high to employ Pion EOS, according to Fig.~1.

\begin{figure}
\includegraphics[scale=0.8]{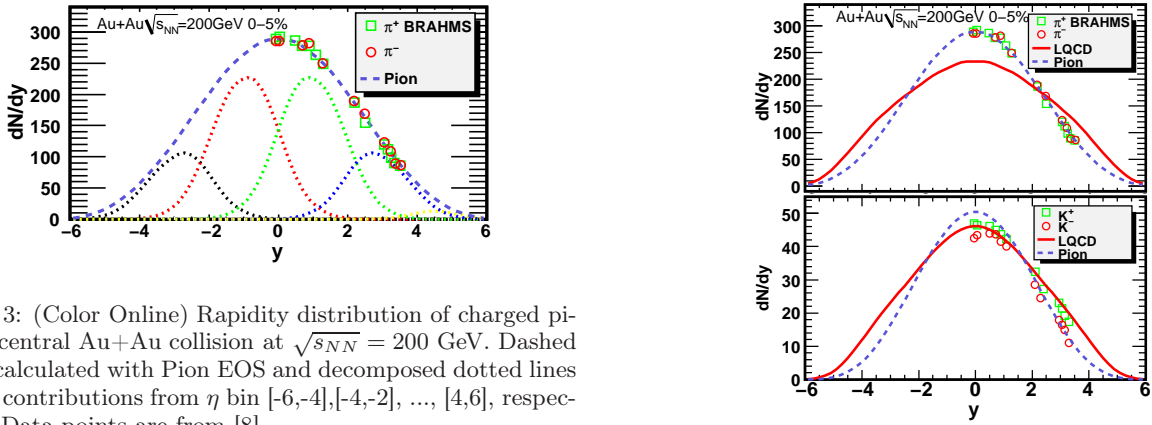}

\caption{\label{fig:Fig3} (Color Online) Rapidity distribution of charged
pions in central Au+Au collision at $\sqrt{s_{NN}}=200$~GeV. Dashed
line is calculated with Pion EOS and decomposed dotted lines for the
contributions from $\eta$ bin [-6,-4],[-4,-2], ..., [4,6],
respectively. Data points are from \cite{BRAHMS}.}
\end{figure}

Then we took LQCD EOS instead. This decreases the temperature at 
$\tau_{{\rm FO}}=2a$ quite a lot, c.f. the right panel of Fig.~2. 
In Fig.~4, the resulted rapidity distribution of charged pions and charged Kaons, shown as solid lines,  and Pion EOS results
shown as dashed lines, are compared with BRAHMS data points~\cite{BRAHMS}. 
While Pion EOS results can reproduce BRAHMS measured pion and Kaon rapidity distribution, this realistic LQCD EOS calculation can not!

The reason is clear. At $\tau_{{\rm FO}}=2a$,   hadrons of large rapidity
mainly freeze out from the fluid at large-$\eta$, where energy
density is low and LQCD EOS coincides with Pion EOS.  While at midrapidity, hadrons mainly freeze out from  fluid at  $\eta \sim 0$ where the energy density is high.
Then the temperature interpreted with LQCD EOS is much lower than that with Pion EOS, which makes less hadron emission at midrapidity.
As a result, the rapidity shape with LQCD EOS becomes broader and deviates from the measured data. 
The correct choice of LQCD EOS itself can not reproduce the data, but the deviation can tell us even more.

%In Fig.4, The normalization factors are 0.072(pion), 0.022(Kaon) 
%for Pion EOS, 0.202(pion), 0.106(Kaon) for LQCD EOS.
%
\begin{figure}
\includegraphics[scale=0.65]{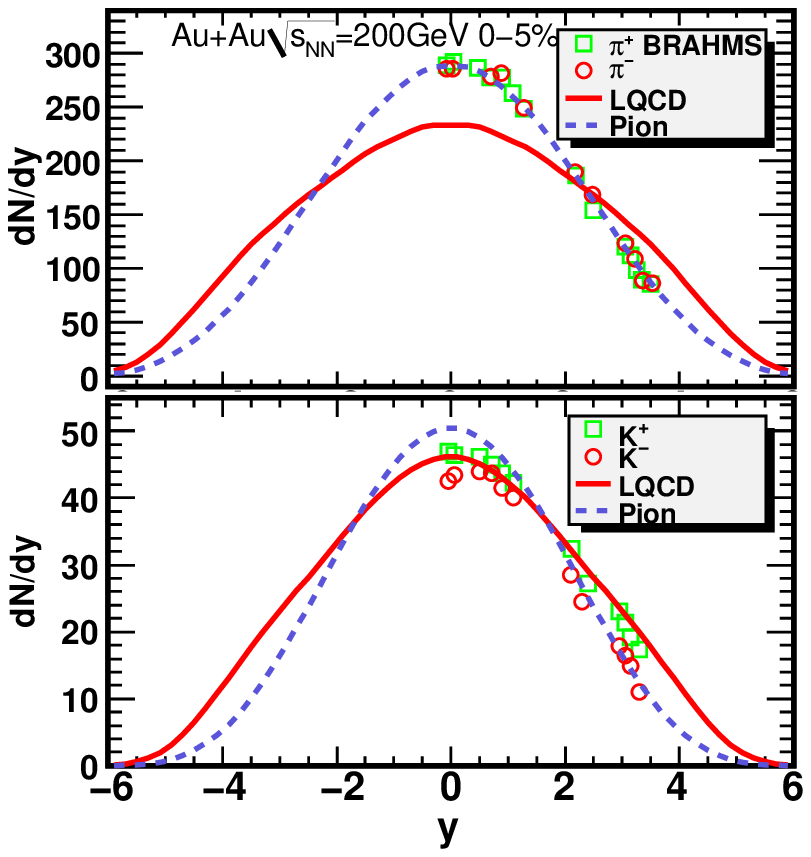}

\caption{\label{fig:Fig4} (Color Online) Rapidity distribution of pion(upper
panel) and Kaon(lower panel) in central Au+Au collision at $\sqrt{s_{NN}}=200$~GeV, with Pion EOS(dashed line) and LQCD EOS(solid line). 
Data points are from \cite{BRAHMS}.}
\end{figure}

To understand the deviation better, we calculated the transverse mass spectra
of charged pions and Kaons from central Au+Au collisions at $\sqrt{s_{NN}}=200$~GeV,
with the FO condition $\tau_{{\rm FO}}=2a$ and two options of EOS.
The result with Pion EOS is shown in Fig.~5, where solid lines for
pions and dashed lines for Kaons, at midrapidity $y=0$ (left panel)
and forward rapidity $y=3.5$ (right panel) and compared with BRAHMS
data where full cycles for $\pi^{-}$ and full squares for $K^{-}$.
The calculation with LQCD EOS is plotted in Fig.~6, where the same
notation has been used. Again, similar results appear at forward rapidity
$y=3.5$ for both EOS, where particles freeze out from quite low energy
density source and two EOS coincidence. 

However, at midrapidity
$y=0$, the inverse slope of the transverse spectra of both pions
and Kaons from LQCD calculation is lower than experimental data, while
the calculation with Pion EOS makes higher the inverse slope for pions,
due to the extremely high temperature interpreted with Pion EOS. 

The second line in lower-right panel of Fig.~2 tells the temperature
interpreted with LQCD is about 200~MeV at $\eta \sim 0$, which is lower 
than Pion EOS value, but still higher than $T_c \sim$ 160~MeV. 
Then Why the calculated inverse slopes of both pions and Kaons are still lower than experimental data? Because there is no transverse expansion in Landau hydrodynamics. 
Hydrodynamical models with transverse expansion and freeze-out at 160~MeV can make higher inverse slope of transverse spectra and reproduce the data, i.e.~\cite{Klaus}.
The deviation tell us, the collective transverse expansion of the collision systems is important to explain the measured transverse spectra.

\begin{figure*}
\includegraphics[scale=0.7]{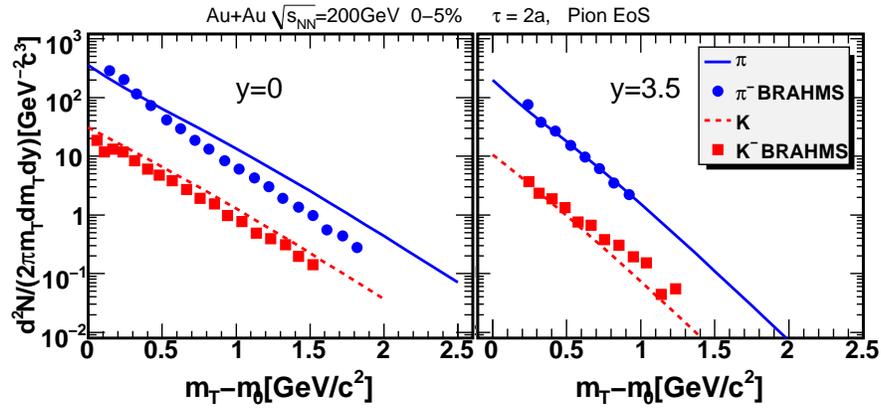}

\caption{\label{fig:Fig5} (Color Online) Transverse mass spectra of charged
pions and Kaons from central Au+Au collisions at $\sqrt{s_{NN}}=200$~GeV
at midrapidity $y=0$ (left panel) and forward rapidity $y=3.5$ (right
panel). The calculation was done with Pion EOS and FO condition $\tau_{{\rm FO}}=2a$,
where solid lines for pions and dashed lines for Kaons. Cycles for
$\pi^{-}$ and squares for $K^{-}$ of BRAHMS data~\cite{BRAHMS}. }
\end{figure*}

\begin{figure*}
\includegraphics[scale=0.7]{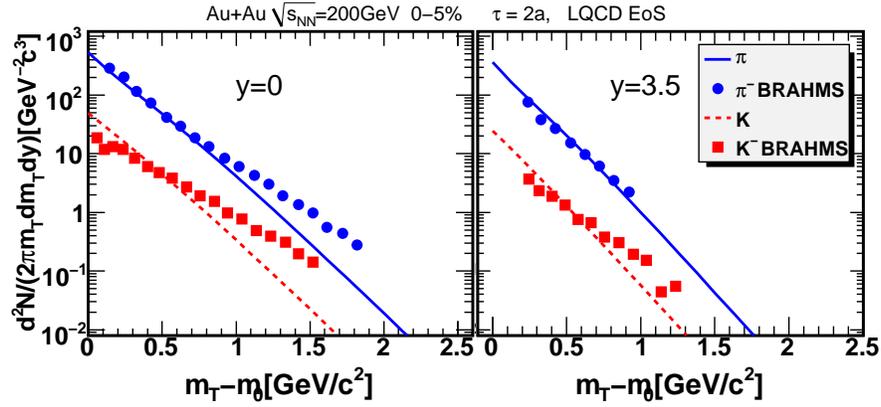}

\caption{\label{fig:Fig6} (Color Online) Same as Fig.~5 except the calculation
was done with LQCD EOS.}
\end{figure*}

The midrapidity transverse momentum spectra of charged pions, Kaons
and antiproton from central Pb+Pb collisions at $\sqrt{s_{NN}}=2.76$~TeV,
with the FO condition $\tau_{{\rm FO}}=2a$ and two options of EOS
are shown in Fig.~7, where solid lines for Pion EOS and dashed lines
for LQCD EOS, and ALICE data~\cite{ALICE} shown as full
cycles. Again we can see Pion EOS gives high inverse slopes than LQCD.
The calculated pion slope with LQCD EOS seems to reproduce the data,
again by accidence. Because the FO condition $\tau_{{\rm FO}}=2a$ makes
a much higher energy density and the temperature at freeze-out is interppreted 
as high as 400~MeV, c.f. the second line in the upper panel of Fig.~2. This
compensates the missing of transverse expansion
in Landau fluid. 

\begin{figure*}
\includegraphics[scale=0.8]{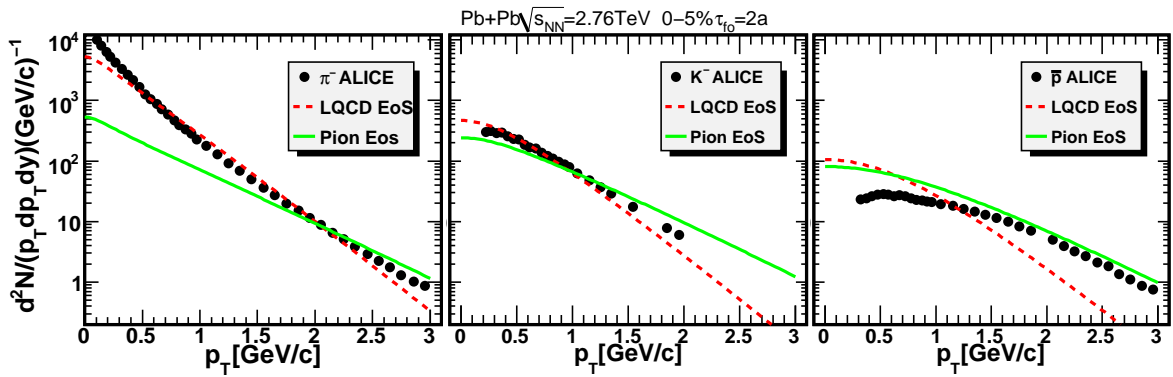}

\caption{\label{fig:Fig7} (Color Online) Transverse momentum spectra of pion
,Kaon and anti-proton from central Pb+Pb collisions at $\sqrt{s_{NN}}=2.76$~TeV.
FO condition $\tau_{{\rm FO}}=2a$, solid lines for Pion EOS and dashed
lines for LQCD EOS. Data points from ALICE collaboration~\cite{ALICE}.}
\end{figure*}

In Fig.~\ref{fig:Fig7}, the deviation of transverse spectra between data and LQCD results, becomes bigger and bigger, as the mass ordering from pion, Kaon to antiproton. This is because, transverse flow drives heavier particles to move along radial direction at the same collective velocity as light particles. However, heavier particles gain more energies and higher inverse slope of the transverse momentum spectra. In another word, transverse expansion of the collision systems becomes more and more important with the increase of hadron mass.

%And the experimental results are from Ref.{[}6]. The normalization factors are 0.008(pion), 0.0092(Kaon), 0.0092(anti-proton) for Pion EOS, and 0.16(pion), 0.052(Kaon), 0.052(anti-proton) for LQCD EOS.

%

As mentioned above, the FO condition $\tau_{{\rm FO}}=2a$ is not
realistic, because the  energy density and temperature  
vary a lot along the beam direction. 
In the following, we take a constant energy density $\epsilon_{{\rm FO}}$ as a FO condition.

In Fig.~8 are shown the transverse mass spectra of pion (upper panel)
and Kaon (lower panel) in central Au+Au collision at $\sqrt{s_{NN}}=200$~GeV,
where left panel for y=0 and right panel for y=3.5, with solid line,
dashed line and dotted line for $\epsilon_{{\rm FO}}$=0.08, 0.16,
0.24~GeV/fm$^{3}$, respectively. The higher $\epsilon_{{\rm FO}}$
can slightly increase the inverse slope of the transverse spectra,
while the slopes are still far from the BRAHMS data, due the missing
of transverse flow in Laudau fluid.

\begin{figure*}
\includegraphics[scale=0.8]{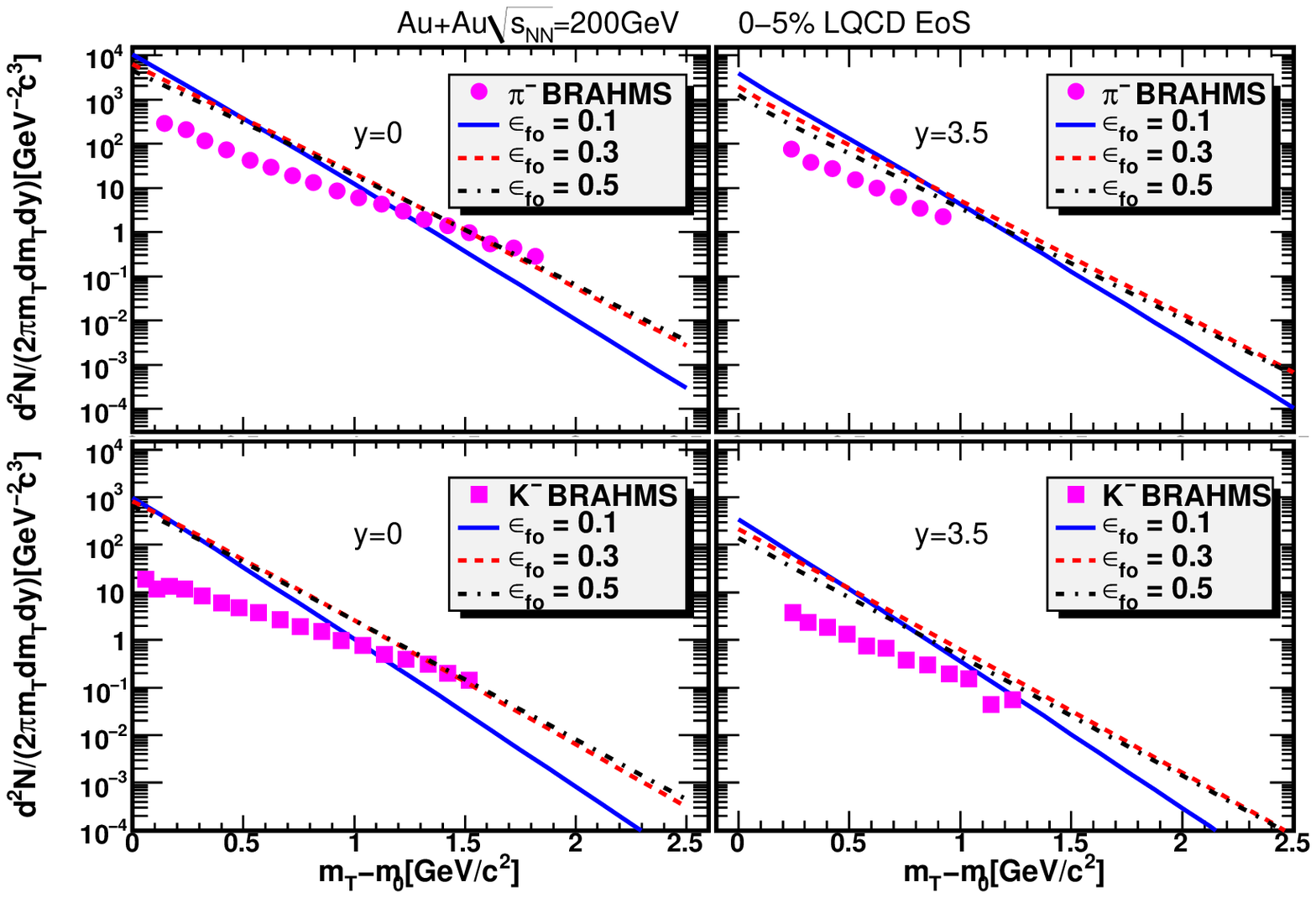}

\caption{\label{fig:Fig8} (Color Online) The transverse mass spectra of pion
(upper panel) and Kaon (lower panel) in central Au+Au collision at
$\sqrt{s_{NN}}=200$~GeV, where left panel for y=0 and right panel
for y=3.5. Solid lines, dashed lines and dotted lines for FO condition
$\epsilon_{{\rm FO}}$=0.08, 0.16, 0.24~GeV/fm$^{3}$, respectively.
Data points are from the BRAHMS collaboration~\cite{BRAHMS}. }
\end{figure*}

In Fig.~9 are shown the midrapidity transverse momentum spectra of
charged pions, Kaons and antiproton from central Pb+Pb collisions
at $\sqrt{s_{NN}}=2.76$~TeV, with the FO condition $\epsilon_{{\rm FO}}$=0.08,
0.16, 0.24~GeV/fm$^{3}$ shown as solid lines, dashed lines and dotted
lines respectively. ALICE data~\cite{ALICE} are shown as full
cycles. Again, higher $\epsilon_{{\rm FO}}$ makes slightly higher
inverse slope. But compared with Fig.~8 for the case of central Au+Au
collision at $\sqrt{s_{NN}}=200$~GeV, the miss of transverse expansion
of Landau fluid in central Pb+Pb collisions at $\sqrt{s_{NN}}=2.76$~TeV
is even stronger, for the larger and hotter collision system.

\begin{figure*}
\includegraphics[scale=0.8]{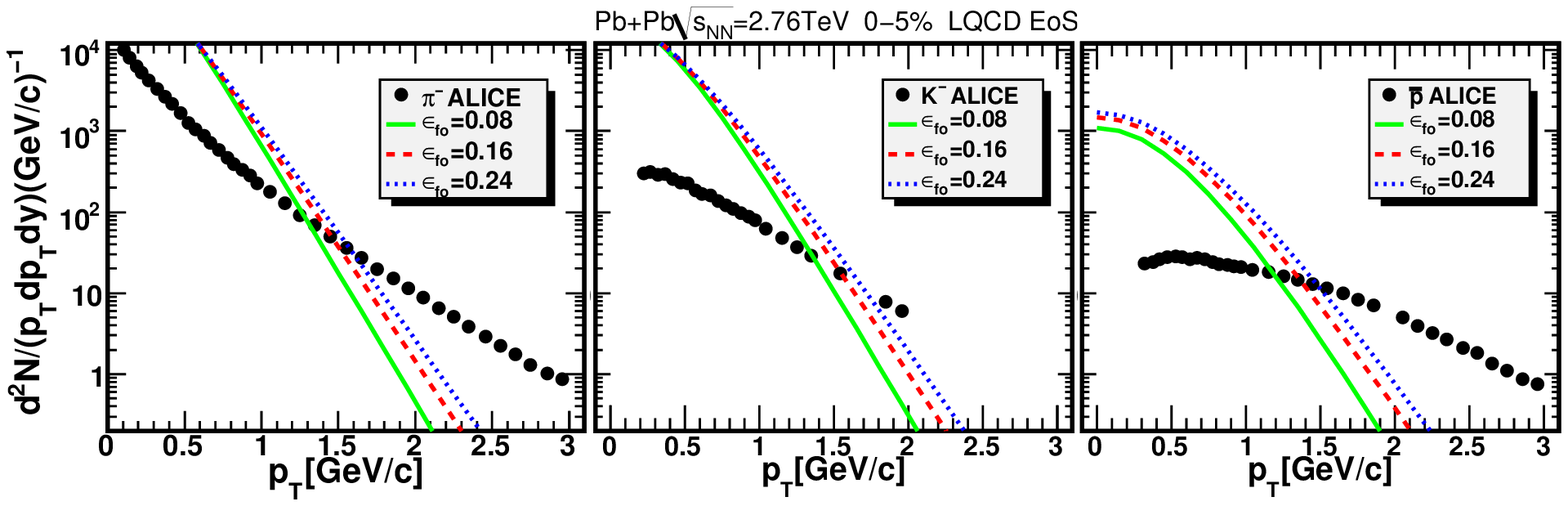}

\caption{\label{fig:Fig10} (Color Online) Transverse momentum spectra of
pion ,Kaon and anti-proton from central Pb+Pb collisions at $\sqrt{s_{NN}}=2.76$~TeV.
Solid lines, dashed lines and dotted lines for FO condition $\epsilon_{{\rm FO}}$=0.08,
0.16, 0.24~GeV/fm$^{3}$, respectively. Data points from ALICE collaboration~\cite{ALICE}. }
\end{figure*}

\section{Conclusions}

In summary, we demonstrated the production of most copious particles from
central Au+Au collisions at $\sqrt{s_{NN}}=200$~GeV and central
Pb+Pb collisions at $\sqrt{s_{NN}}=2.76$~TeV in the framework of
Landau hydrodynamical model. 

The measured rapidity distribution of pions and kaons from
central Au+Au collisions at $\sqrt{s_{NN}}=200$~GeV and the transverse momentum spectrum of pions from Central Pb+Pb collisions at $\sqrt{s_{NN}}=2.76$~TeV
can be reproduced with Landau model by accidence, due to the choice of FO condition $\tau_{{\rm FO}}=2a$ and the employment of Pion EOS. This FO condition makes most hadrons freeze out from a fluid at very high energy density. The employment of Pion EOS causes that the temperature at which hadrons are emitted is interppreted to be as high as 400~MeV.
The emission of hadrons at high density and high temperature compensates the missing of transverse expansion in Landau hydrodynamical model. 

With a more realistic choice of FO condition, fixed at a relatively low  energy density, and the employment of LQCD EOS for particle production, we can see that transverse expansion becomes important to explain bulk hadron production in high energy nuclear collisions. Especially, with the increase of collision energy from RHIC to LHC, and with the increase of the mass of produced hadrons, transverse expansion becomes more and more important.

\begin{acknowledgments}
This work is supported by the Natural Science Foundation of China
under the project No.~10975059 and No.~11275081. 
\end{acknowledgments}


\begin{thebibliography}{10}
\bibitem{Fermi} E.~Fermi, Progr. Theo. phys. \textbf{5} (1951)570;
Phys.Rev. \textbf{81} (1951) 683.


\bibitem{Aoki:2009sc} 
  Y.~Aoki, S.~Borsanyi, S.~Durr, Z.~Fodor, S.~D.~Katz, S.~Krieg and K.~K.~Szabo,
  %``The QCD transition temperature: results with physical masses in the continuum limit II.,''
  JHEP {\bf 0906}, 088 (2009)
  [arXiv:0903.4155 [hep-lat]].
  %%CITATION = ARXIV:0903.4155;%%


\bibitem{Landau} L. D. Landau, Izv. Akad. Nauk. SSSR \textbf{17},
51 (1953);  S. Z. Belenkij and L. D. Landau, Usp. Fiz. Nauk
\textbf{56}, 309 (1955); Nuovo Cimento Supp1. \textbf{3}, 15 (1956).



\bibitem{Wong} C.Y.Wong, Phys. Rev.C \textbf{78}, 054902 (2008).

\bibitem{Csernai}  M.~Zetenyi and L.~P.~Csernai,
  %``Studying freeze-out and hadronization in the Landau hydrodynamical model,''
  Phys.\ Rev.\ C {\bf 81}, 044908 (2010)
  [arXiv:1003.3757 [nucl-th]].
  %%CITATION = ARXIV:1003.3757;%%

\bibitem{Cooper_Frye} F.Cooper and G.Frye, Phys. Rev.D \textbf{10},
186 (1974).


\bibitem{Aoki:2005vt} 
  Y.~Aoki, Z.~Fodor, S.~D.~Katz and K.~K.~Szabo,
  %``The Equation of state in lattice QCD: With physical quark masses towards the continuum limit,''
  JHEP {\bf 0601}, 089 (2006)
  [hep-lat/0510084].
  %%CITATION = HEP-LAT/0510084;%%

\bibitem{BRAHMS} I. G. Bearden \textit{et al.} (BRAHMS collaboration),
Phys. Rev. Lett. \textbf{94}, 162301 (2005).

\bibitem{Klaus} K.~Werner, I.~.Karpenko, T.~Pierog, M.~Bleicher and K.~Mikhailov,
  %``Event-by-Event Simulation of the Three-Dimensional Hydrodynamic Evolution from Flux Tube Initial Conditions in Ultrarelativistic Heavy Ion Collisions,''
  Phys.\ Rev.\ C {\bf 82}, 044904 (2010)
  [arXiv:1004.0805 [nucl-th]].
  %%CITATION = ARXIV:1004.0805;%%

\bibitem{ALICE} M. Floris (CERN), J. Phys. G \textbf{G38}, 124025(2011). 
\end{thebibliography}
\end{document}